\documentclass[prl,twocolumn, showpacs,preprintnumbers,amsmath,amssymb,superscriptaddress,floatfix]{revtex4}

\usepackage[]{graphicx}      
\usepackage{bm}             	
\usepackage{times}			
\usepackage{tabularx}

\newcommand{\IEF}{Institut d'Electronique Fondamentale, CNRS, UMR 8622, 91405 Orsay, France}
\newcommand{\UPS}{Universit{\'e} Paris-Sud, UMR 8622, 91405 Orsay, France}
\newcommand{\IMEC}{IMEC, Kapeldreef 75, B-3001 Leuven, Belgium}
\newcommand{\USFD}{Department of Engineering Materials, University of Sheffield, Sheffield S1 4DU, United Kingdom}
\newcommand{\KUL}{Laboratorium voor Vaste-Stoffysica en Magnetisme, K. U. Leuven, Celestijnenlaan 200 D, B-3001 Leuven, Belgium}
\newcommand{\StP}{Saint P{\"o}lten University of Applied Sciences, Matthias Corvinus Str. 15, St. P{\"o}lten A-3100, Austria}

\begin{document}

%
%
\title{Agility of vortex-based nanocontact spin torque oscillators}

\author{M. Manfrini}
\affiliation{\IMEC}
\affiliation{\KUL}
\author{T. Devolder}
\email{thibaut.devolder@u-psud.fr}
\author{Joo-Von Kim}
\affiliation{\IEF}
\affiliation{\UPS}
\author{P. Crozat}
\author{N. Zerounian}
\affiliation{\UPS}
\affiliation{\IEF}
\author{C. Chappert}
\affiliation{\IEF}
\affiliation{\UPS}
\author{W. Van Roy}
\author{L. Lagae}
\affiliation{\IMEC}
\affiliation{\KUL}
\author{G. Hrkac}
\affiliation{\USFD}
\author{T. Schrefl}
\affiliation{\USFD}
\affiliation{\StP}

\date{\today}                                           
%
%
\begin{abstract}
We study the agility of current-tunable oscillators based on a magnetic vortex orbiting around a point contact in spin-valves. Theory predicts frequency-tuning by currents occurs at constant orbital radius, so an exceptional agility is anticipated. To test this, we have inserted an oscillator in a microwave interferometer to apply abrupt current variations while time resolving its emission. Using frequency shift keying, we show that the oscillator can switch between two stabilized frequencies differing by 25\% in less than ten periods. With a wide frequency tunability and a good agility, such oscillators possess desirable figures of merit for modulation-based rf applications.
\end{abstract}


\maketitle

%
%

In the context of spin torque oscillators, metallic nanocontacts on spin-valves represent an important system in which both spin wave radiation~\cite{Rippard:PRL:2004, Rippard:PRL:2005, Pufall:APL:2005, Kaka:Nature:2005, Mancoff:Nature:2005} and large-amplitude excitations of magnetic vortices~\cite{Pufall:PRB:2007, Mistral:PRL:2008, Devolder:APL:2009, Ruotolo:NN:2009} are possible. Indeed for certain nanocontact sizes and multilayer configurations, the Oersted field associated with the current can nucleate a magnetic vortex, which is subsequently set into steady-state rotation about the nanocontact by competing damping and spin-torques~\cite{Pufall:PRB:2007,Mistral:PRL:2008}. A stable oscillation of the point contact resistance is generated at 100-500 MHz as the vortex orbits around the magneto-resistive contact, which opens up possible applications as compact rf oscillators~\cite{Keller:APL:2009} operating at zero field~\cite{Devolder:APL:2009} and allowing multi-octave frequency coverage. From an rf application perspective, the frequency tunability of such vortex based spin torque oscillators is outstanding since the free running frequency can be varied by a factor of more than three~\cite{Devolder:APL:2009}. However many rf applications require modulation schemes. The oscillator agility -- the rate at which the frequency can be effectively tuned -- is therefore an essential figure of merit that needs to be measured and understood. 

From prior studies, a qualitative understanding of the nature of vortex-based nanocontact oscillator has already been achieved using a rigid-vortex model in which the translational dynamics are studied after integrating out the internal degrees of freedom~\cite{Mistral:PRL:2008}. A major success of the model is the prediction that the vortex orbital frequency $F$ is linear with the applied current magnitude $I_{dc}$, $F \propto {\alpha I_{dc}} / {\sigma a^2}$, where $\alpha$ is the Gilbert damping constant, $\sigma$ represents a spin transfer efficiency and $a$ is the point contact radius. Another, less emphasized prediction of the model is that the radius $R$ of the vortex orbit should vary as $R \propto {\sigma a^2} / \alpha $, which is \textit{independent} of the current as long as micromagnetics permits the existence of the vortex. An important consequence of these two predictions is that an abrupt change in the current is expected result in an abrupt change in the orbital frequency $F$, while leaving the vortex orbital radius $R$ unchanged. This suggests that an infinitely-agile oscillator is possible based on such vortex oscillations, which would be of considerable interest for radiofrequency applications.

\begin{figure}
\includegraphics[width=7cm]{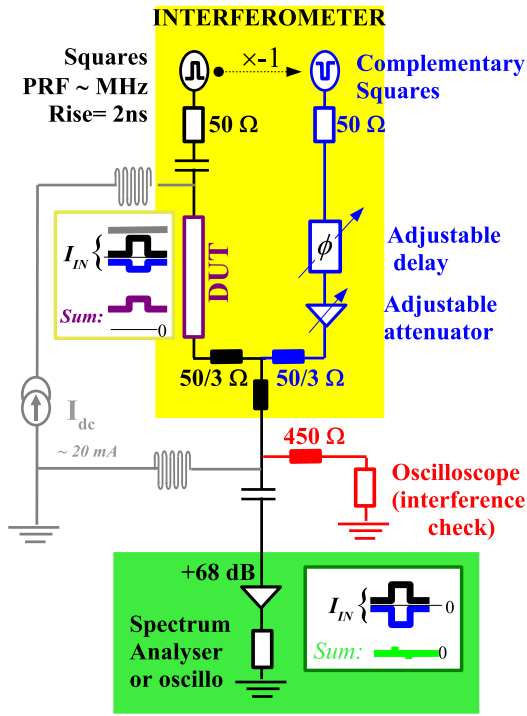}
\caption{\label{fig:1}(Color online) Microwave interferometer design to characterize the agility of the vortex-based spin torque oscillators. The current wave effectively applied to the device is sketched in the yellow box on the middle left. A sketch of current wave arriving at the amplifiers is inserted at the bottom right.}
\end{figure}

\begin{figure}
\includegraphics[width=9cm]{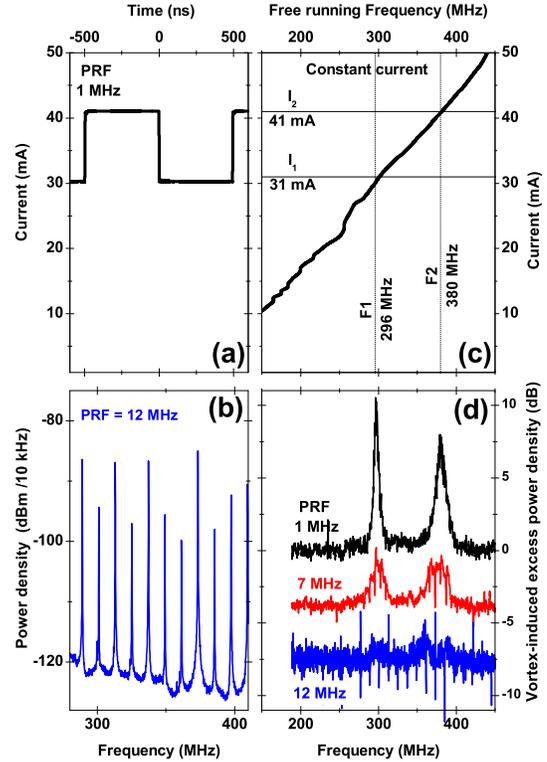}
\caption{\label{fig:2}(Color online) (a) Current waveform applied to the device (PRF=1 MHz). (b) Noise power density at the output of the interferometer for a PRF of 12 MHz when no vortex is present. (c) Frequency dependence versus current in constant current mode (PRF=0). (d) Additional Noise power density when a vortex has been priorly nucleated using the dc current branch. For clarity, the curves for 7 and 12 MHz PRF have been shifted vertically by -4 and -8 dB.}
\end{figure}

\begin{figure}
\includegraphics[width=9cm]{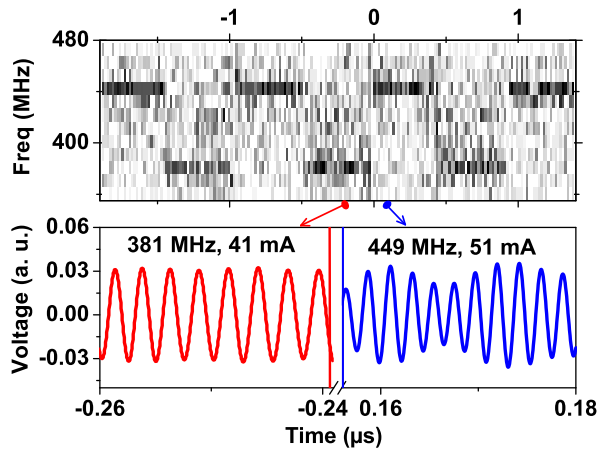}
\caption{\label{fig:3}(Color online) (a) Time-localized power spectra of the vortex signal calculated for a current wave of 1 MHz PRF swapping between 41 and 51 mA. The pixel size is 10 ns $\times$ 10 MHz and its color scales with the power density. (b) Zoom on single-shot voltage waveforms before and after a current command transition. The waveforms have been band-pass filtered from 350 to 480 MHz. The apparent amplitude noise is an experimental artifact.}
\end{figure}

In this article, we present an experimental study of the agility of vortex-based nanocontact spin torque oscillators. We have implemented a novel microwave interferometry experiment (Fig.~\ref{fig:1}) that allows us to apply abrupt current changes to the oscillator while maintaining the possibility to frequency-resolve (Fig.~\ref{fig:2}) and time-resolve (Fig.~\ref{fig:3}) its rf emission. We show that the vortex oscillator can switch between frequencies within a few oscillation periods, and that the oscillation is quickly stabilized, confirming the exceptional agility expected from theory. As an illustration of this agility, we implement a frequency shift keying (FSK) modulation scheme and switch between two carriers at 380 and 450 MHz every 40 ns, a performance that is hard reach with standard voltage controlled oscillators~\cite{Grebennikov:book:2007}.
 
Our point contacts are fabricated on top of spin-valves of composition IrMn(6) / Co$_{90}$Fe$_{10}$(4.5) / Cu(3.5) / Co$_{90}$Fe$_{10}$(1.5) / Ni$_{80}$Fe$_{20}$(2), with a nominal contact diameter $2a=100$ nm. The figures in parentheses denote the film thicknesses in nm. Details of the fabrication process~\cite{Mistral:PRL:2008} and of the properties of a generic oscillator~\cite{Devolder:APL:2009} have been given elsewhere. Positive currents correspond to electrons passing from the free to the pinned layer. For the representative device under test (DUT) studied in detail here, vortex nucleation at 50 mA and subsequent vortex motion down to 9 mA are observed only for this polarity. Its emission frequency is linear with the applied current (Fig.~\ref{fig:2}c), with a slope of 7.4 MHz/mA and a zero current extrapolated intercept of 77 MHz. The device and its series electrodes have a resistance of $R_P=7.46 \;\Omega$ and a giant magnetoresistance (GMR) of $\Delta R= 25 \;\textrm{m}\Omega$. 

%
%

The device is inserted in a microwave interferometer with two balanced arms (Fig.~\ref{fig:1}) whose role is to allow the weak oscillator signal to be measured while avoiding to be blinded by the high amplitude fast current modulation.  A dc current branch is added to the interferometer for controlled vortex nucleation and annihilation. The working principle of the interferometer is the following. Both arms are fed by square waves, with a pulse repetition frequency (PRF) in the MHz range and 2 ns transition times (Fig. \ref{fig:2}a). The square wave of the right arm of the interferometer is the exact complement of that of the left arm. The device is placed in series with the left arm and is subject to the sum of dc current, a direct left-coming wave, and an attenuated wave coming from the right arm, all this resulting in a square wave with dc offset. At the device, the vortex oscillation leads to a GMR variation that manifests itself as a small ripple superimposed on the plateaus of this square wave (not shown in Fig.~\ref{fig:1}). A power combiner consisting of a three $50/3 \;\Omega$ resistance bridge adds the left-coming square wave to its right-coming complement, leaving only the signatures of the vortex oscillations, in principle, which can then be amplified and analyzed. The destructive interference of the square waves is optimized using an adjustable delay line and an adjustable attenuator. The cancellation quality is checked on a fraction of the signal coupled to an oscilloscope using a $450 \;\Omega$ pick-off tee. The square wave cancellation is perfect except during the rising and falling edges of the square waves because they jitter by typically 2 ns due to instrumental limitations. This saturates the amplifiers after each current transition; their recovery takes some 25 ns. In addition to the vortex signals, this yields a time-periodic pulse pattern, that when translated in frequency domain, is a frequency comb with a peak spacing equal to PRF and a roll-off extending to typically 500 MHz (Fig. \ref{fig:2}(b)). When using the spectrum analyzer, maintaining the ability to detect the vortex signal despite this intense frequency comb requires to use a resolution bandwidth (RBW) far below the PRF. An RBW of 10 kHz is for instance chosen in Fig.~\ref{fig:2}b.

%

Using this setup, we have studied the response of the oscillator to a varying external stimulus. We have submitted the device to a square wave current modulation between 31 and 41 mA (Fig.~\ref{fig:2}a). In a constant current experiment, these currents would yield frequencies of $F_1=296$ and $F_2=380$ MHz, respectively (Fig.~\ref{fig:2}c). These two current values are below the nucleation threshold, so that the measurement can be done with (Fig.~\ref{fig:2}b) or without (not shown) vortex, and the vortex-induced signal can be isolated by a simple subtraction (Fig.~\ref{fig:2}d). At low PRF (below 1 MHz), the device toggles between modes 1 and 2, replicating the frequencies, linewidth and power of the corresponding dc experiment. Note that the upper frequency mode ($f_2$) has a larger linewidth $\Delta f_2$ than the lower frequency mode. When increasing the PRF to $7 \textrm{MHz}$, $F_1$,  $F_2$ and $\Delta F_2$ remain unaffected.  In contrast, the linewidth of mode 1 increases and starts to be limited by (2PRF)$^{-1}$, i.e. the time interval during which the required current for this mode is applied. This linewidth increase indicates that phase coherence is completely lost in between two repeats of the first mode. When further increasing the PRF to 12 MHz, each current is applied less than 42 ns. This corresponds to some 20 periods from which half of them can not be measured because of transient saturations of the amplifiers at the transition. Despite this, two separated modes still emerge faintly from the noise level at frequencies close to $F_1$ and $F_2$. These results allow us to conclude that the time needed for the frequency to change and stabilize is below an upper bound of 20 ns.

%

To characterize the abruptness of the frequency transition and the time required for the emission frequency to be stabilized, we employed a fast single-shot oscilloscope in place of the spectrum analyzer and subtracted the signal obtained when no vortex had been nucleated. Before and long after the transition, the mode frequency can be directly visualized using the time-resolved voltage signal (Fig. \ref{fig:3}b). However, direct time-domain visualization of the transition is not possible because of transient saturation and gradual recovery of the amplifiers. In order to better visualize the transition, we have segmented the time-domain data into time intervals of 100 ns on which we can calculate the power spectra. Our frequency resolution is therefore 10 MHz, i.e. sufficient to discriminate between $F1$ and $F_2$. Each 100 ns window is slid in steps of 10 ns, providing a large overlap between successive segments. Our time resolution is therefore essentially 10 ns with a convolution with a 100 ns aperture, such that 2-3 pixels after each current transition are invalidated by the amplifier transient saturation. The two modes appear as perfectly horizontal dark bars in Fig. \ref{fig:3}a, indicating that whenever a mode is present its frequency is stable. In addition, the frequency variations in Fig. \ref{fig:3}a clearly indicate that the spacing between the bars is typically 2-3 pixels (20-30 ns), i.e. our time resolution limit, despite the reduced signal to noise ratio. We conclude that the frequency transition is not gradual but abrupt, at least down to the frequency resolution of our experiment.

In conclusion, we have studied the agility of a vortex nanocontact spin torque oscillator using a novel microwave interferometer technique. Using a combination of time and frequency domain measurements, we show that the oscillator can switch between two stabilized frequencies under an upper bound of 20 ns. A wide frequency tunability and a very good agility is a unique feature of vortex oscillators that makes them appropriate for rf modulation schemes, which have much potential for applications.

This work has been supported by the Triangle de la Physique contract 2007-051T. M. M. is supported by the European Community under the 6th FP for the Marie Curie RTN SPINSWITCH, contract n$^{\circ}$ MRTN-CT-2006-035327.




\end{document}